# Magnetic imaging and statistical analysis of the metamagnetic phase transition of FeRh with electron spins in diamond

Guillermo Nava Antonio[1,2*], Iacopo Bertelli[1,3*], Brecht G. Simon[1], Rajasekhar Medapalli[4,5], Dmytro Afanasiev[1,6], and Toeno van der Sar[1†]

[1] Department of Quantum Nanoscience, Kavli Institute of Nanoscience, Delft University of Technology, 2628 CJ Delft, The Netherlands.
[2] Current address: Cavendish Laboratory, University of Cambridge, CB3 0HE Cambridge, United Kingdom
[3] Huygens-Kamerlingh Onnes Laboratorium, Leiden University, 2300 RA Leiden, The Netherlands.
[4] Center for Memory and Recording Research, University of California, San Diego, 92093-0401 California, USA
[5] Current address: Department of Physics, Lancaster University, LA1 4YB Lancaster, United Kingdom
[6] Current address: Department of Physics, University of Regensburg, 93053 Regensburg, Germany

[*] These authors contributed equally to this work.
[†] Corresponding author. E-mail: t.vandersar@tudelft.nl

Dated: 27 March 2021

**Abstract**

Magnetic imaging based on nitrogen-vacancy (NV) centers in diamond has emerged as a powerful tool for probing magnetic phenomena in fields ranging from biology to physics. A key strength of NV sensing is its local-probe nature, enabling high-resolution spatial images of magnetic stray fields emanating from a sample. However, this local character can also form a drawback for analysing the global properties of a system, such as a phase transition temperature. Here, we address this challenge by using statistical analyses of magnetic-field maps to characterize the first-order temperature-driven metamagnetic phase transition from the antiferromagnetic to the ferromagnetic state in FeRh. After imaging the phase transition and identifying the regimes of nucleation, growth, and coalescence of ferromagnetic domains, we statistically characterize the spatial magnetic-field maps to extract the transition temperature and thermal hysteresis width. By analysing the spatial correlations of the maps and their dependence on an external magnetic field, we investigate the magnetocrystalline anisotropy and detect a reorientation of domain walls across the phase transition. The employed statistical approach can be extended to the study of other magnetic phenomena with NV magnetometry or other sensing techniques.

Magnetic imaging and statistical analysis of the metamagnetic phase transition of FeRh with electron spins in diamond

1. Introduction

Nitrogen-vacancy (NV) centers are point-like defects in the diamond lattice hosting an electron spin that can be used as a sensor of magnetic fields[1]. In the last decade, NV centers have been developed into powerful tools for probing magnetic phenomena in both condensed-matter and biological systems[2,3]. Key strengths of the technique are its nanoscale spatial resolution[4], high magnetic field sensitivity[5], and its operability under cryogenic conditions to above room temperature[6]. However, its local-probe nature can also form a challenge for extracting quantities related to the macroscopic properties of a system such as its phase transition temperature and hysteresis width. In addition, while real-space images provide a powerful visualization of spatial textures such as domain structures, extracting the temperature- or field-driven evolution of important parameters such as domain sizes requires defining suitable figures of merit.

In this work, we image the magnetic stray fields generated by an FeRh thin film using an ensemble of NV centers in a diamond chip (Fig. 1a) and employ statistical methods[7] to characterize its temperature-driven metamagnetic phase transition (MMPT)[8]. By analyzing the spatial variations and correlations of measured stray fields as a function of temperature, we extract the transition temperature and hysteresis width of the phase transition. Furthermore, we investigate the magnetocrystalline anisotropy of the sample, identifying a reorientation of ferromagnetic (FM) domain walls during the MMPT triggered by the application of a bias field along a magnetic hard axis. Our results show that combining real-space imaging with statistical figures of merit enables a better understanding of how macroscopic properties arise from the local state of a system.

The alloy FeRh undergoes a first-order phase transition from an antiferromagnetic (AFM) to a FM state when increasing its temperature (Fig. 1b) above a transition temperature ($T_t \approx 370$ K) that depends on the strain[9], bias field[10], and stoichiometry of the sample[11]. This MMPT is accompanied by a volume expansion of $\sim 1\%$ and an abrupt decrease in the resistivity[12], as well as an interplay between AFM and FM domains that was recently shown to lead to an intricate domain structure[13]. As such, FeRh constitutes a platform that is well suited for investigating the interaction between magnetic, electronic, and structural degrees of freedom. The MMPT of FeRh is also of practical interest due to its potential for data storage technologies, such as AFM resistive memories[14], heat-assisted magnetic recording[15], and low-power spintronic devices[16].

2. Results

*2.1 NV imaging of the MMPT in FeRh*

Our sensing platform is based on an NV-containing diamond chip positioned on top of a 20-nm-thick epitaxial FeRh film that is grown on MgO (Fig. 1a) by DC magnetron sputtering[17]. As shown by vibrating sample magnetometry, the transition temperature of the film is ~370 K with a hysteresis width of approximately 20 K under an in-plane bias field of 150 mT (Fig. 1b). The temperature-dependent changes of the FeRh magnetization cause variations of the stray magnetic field above the film, which, in turn, induce Zeeman shifts in the electron spin resonance

Magnetic imaging and statistical analysis of the metamagnetic phase transition of FeRh with electron spins in diamond

(ESR) frequencies of the NV centers[1]. We polarize the NV spins by optical pumping and detect ESR transitions between their $|0\rangle$ and $|\pm1\rangle$ spin sublevels by applying a microwave (MW) field using a nearby bonding wire and monitoring the spin-dependent photoluminescence (Fig. 1c). A DC bias field $\mathbf{B_0}$ is applied along one of the four possible NV orientations (i.e., the $\langle 111 \rangle$ directions of the diamond lattice) to select two ESR transitions among the eight possible ones and to magnetize the FeRh film. Specifically, the bias field is oriented at ~35° out of plane and has an in-plane component $\mathbf{B_0^\parallel}$ that we orient alternatively along a magnetic easy or hard axis of FeRh. The out-of-plane magnetization of FeRh induced by the field is negligible since the shape anisotropy of the film favors in-plane magnetization and the bias fields used in this work are much smaller than the saturation magnetization ($\mu_0 M_s \approx 1.5$ T[18]). This, together with the magnetocrystalline anisotropy of FeRh, results in four in-plane $\langle 100 \rangle$ easy axes[19].

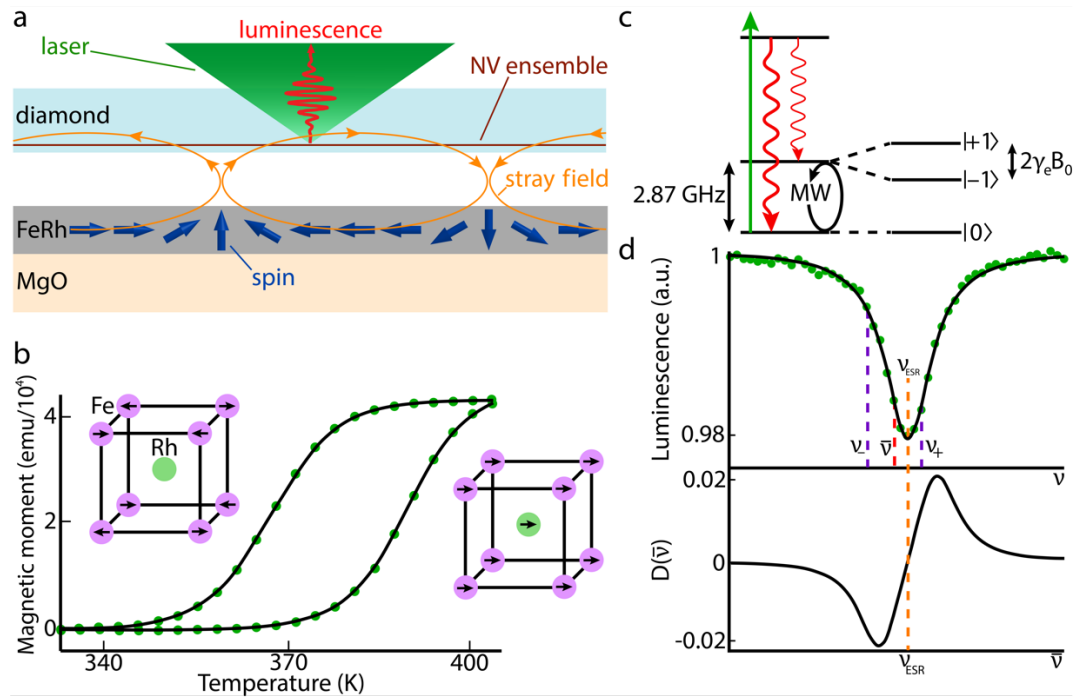

**Figure 1. NV detection of the FeRh stray field. (a)** Schematic of the experimental setup. The stray field of the FeRh film is detected by the NV sensing layer in a diamond chip. Scanning the laser allows to spatially map the FeRh stray field. **(b)** Temperature dependence of the magnetic moment of the 20-nm-thick FeRh film measured by vibrating sample magnetometry under an in-plane bias field of 150 mT, parallel to the [100] FeRh crystalline direction (easy axis). Below the transition temperature $T_t \approx 370$ K, the Fe spins are ordered in an AFM configuration and the Rh atoms carry no net magnetic moment (left inset). Above $T_t$, the Rh atoms gain a finite moment and align ferromagnetically parallel to the Fe spins (right inset)[8]. **(c)** Simplified energy level diagram of an NV center. The NV is optically excited with a green laser and its spin state monitored via spin-dependent photoluminescence. Microwaves drive ESR transitions between the $m_s = |0\rangle$ and $|\pm1\rangle$ spin states. The magnetic field is extracted by measuring the Zeeman-split ESR frequencies. **(d)** Feedback measurement of NV ESR frequencies. Top: ESR spectrum obtained by sweeping the MW frequency while recording the NV photoluminescence. Bottom: Error function resulting from estimating the ESR frequency with a feedback algorithm. In this protocol, only two MW pulses are applied with frequencies $\nu_\pm$, symmetric around an estimate $\bar{\nu}$ of the ESR frequency $\nu_{ESR}$. The photoluminescence

Magnetic imaging and statistical analysis of the metamagnetic phase transition of FeRh with electron spins in diamond difference at these frequencies, denoted $D(\bar{v})$, is proportional to the error $\bar{v} - v_{ESR}$, which we use as feedback signal to adjust the MW pulse frequencies while scanning.

To determine the magnetic field, we measure the two ESR transition frequencies of the selected NV family (Fig. 1d), which yield the projection $B_{NV}$ of the stray field onto the NV axis[1]. Performing this measurement at each pixel by sweeping the MW frequency while detecting the photoluminescence takes several hours for images of 100×100 pixels. Instead, we employ a feedback scheme[20] in which, at each pixel, the photoluminescence is only measured at two frequencies $v_\pm$ (Fig. 1d), thereby speeding up the data collection by about an order of magnitude.

Using this method, we map the stray field generated by the FeRh thin film across the MMPT, and in this way we image the evolution of its magnetic order, when $\mathbf{B}_0^\parallel$ is applied along a hard axis (Fig. 2a – g). Below the phase transition temperature ($T_t$), we observe a mostly vanishing $B_{NV}$, since AFM and nanometric FM domains generate negligible magnetic fields at the NV location. This is because the stray field of these features decays exponentially with the NV-FeRh distance[2], which is around 1 micrometer in our experiments. The appearance of regions of non-zero field at ~368 K indicates the nucleation of FM domains, which grow and coalesce as the temperature is increased further. The magnetization direction results from a competition between the Zeeman energy and the anisotropy energy, the former favoring the alignment of spins with the bias field and the latter with the in-plane easy axes.

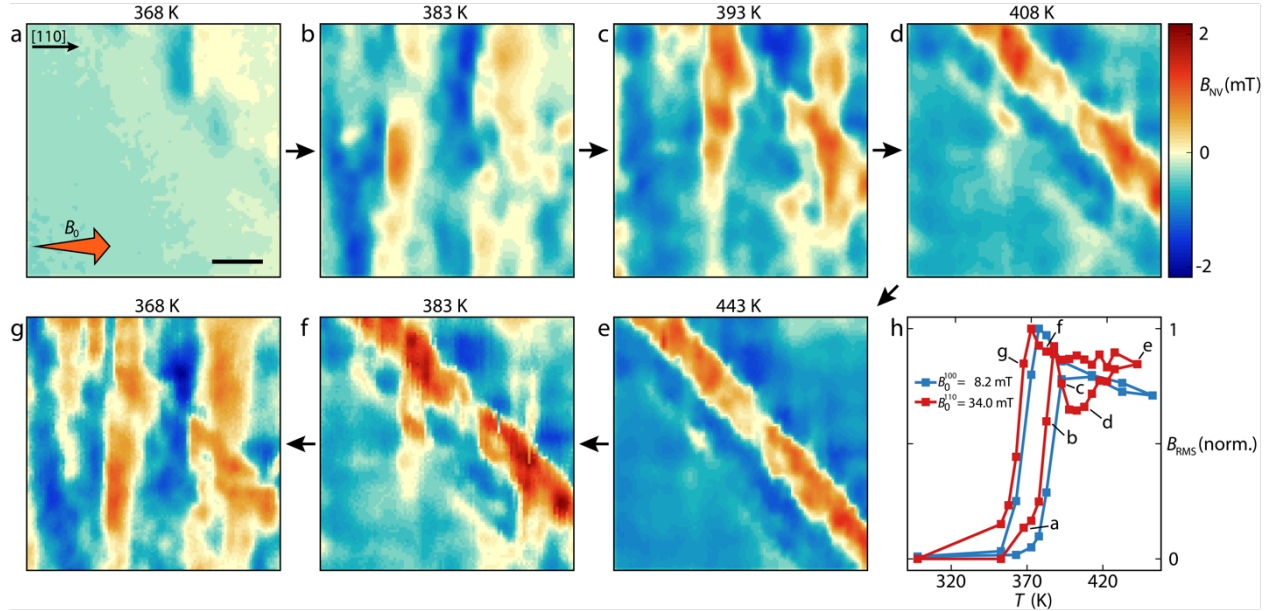

**Figure 2. Imaging the meta-magnetic phase transition (MMPT) of FeRh with NV magnetometry. (a) – (g)** Magnetic-field maps measured across the MMPT after subtraction of the bias field, in the heating (a - e) and cooling (e - g) branches of the experiment. The in-plane component of the bias field is parallel to the [110] FeRh direction. All maps in this work contain 100×100 pixels with a pixel size of 0.5 μm. Scale bar, 10 μm. $B_0$ = 34.0 mT. **(h)** Temperature evolution of the normalized RMS deviation of the measured stray-field maps. The blue (red) line corresponds to the in-plane component of the bias field being oriented along an easy (hard) axis and $B_0$ = 8.2 mT ($B_0$ = 34.0 mT).

Magnetic imaging and statistical analysis of the metamagnetic phase transition of FeRh with electron spins in diamond

Upon decreasing the temperature, we observe thermal hysteresis (i.e., by comparing Fig. 2a and 2g). This hysteretic behavior, together with the coexistence of different magnetic phases represented in Fig. 2a–b, are in line with the well-known first-order character of this phase transition[8].

## 2.2 Quantitative determination of $T_t$ and hysteresis width

To quantitatively analyze the evolution of the stray field and the underlying changes in the magnetization of the FeRh film, we calculate the root-mean-square (RMS) deviation of the measured magnetic-field maps. This figure of merit, which gauges the spatial variations of the field, is computed as[7]

$$B_{\mathrm{RMS}} = \sqrt{\frac{1}{n}\sum_{i=1}^{n}\left(B_{\mathrm{NV},i} - \langle B_{\mathrm{NV}} \rangle\right)^2}, \qquad (1)$$

where $B_{\mathrm{NV},i}$ is the field along the selected NV axis at pixel $i$, $n$ is the total number of pixels, and $\langle B_{\mathrm{NV}} \rangle$ is the average field over the region of interest. The temperature dependence of $B_{\mathrm{RMS}}$ (Fig. 2h) exhibits an abrupt increase at ~373 K, which gives a clear quantitative indication of the transition temperature, unlike the spatial magnetic-field maps discussed previously.

Below $T_t$, $B_{\mathrm{RMS}}$ vanishes as there are no detectable FM domains within the AFM matrix[2]. As the FM domains start to appear, the homogeneity of the stray field is reduced, resulting in an increase of $B_{\mathrm{RMS}}$ between 368 K and 388 K. With $\mathbf{B}_0^{\parallel}$ parallel to an easy axis ([100]), the existing FM regions are expected to be magnetized along this direction and grow in size with increasing temperature, making the $B_{\mathrm{NV}}$ maps more uniform. The observed reduction of $B_{\mathrm{RMS}}$ with further increasing temperatures is expected as a consequence of the magnetization decrease due to thermally excited magnons.

For $\mathbf{B}_0^{\parallel} \parallel [110]$, the anisotropy and in-plane bias fields point along different directions. The magnetization direction of the FM domains results from a balance between these two contributions, which evolve in a non-trivial way as function of the sample temperature, leading to a non-monotonous $T$ dependence of $B_{\mathrm{RMS}}$ above 388 K, which will be further investigated in the next sections. The quantitative character of the applied statistical tools enables detecting small differences between the $\mathbf{B}_0^{\parallel} \parallel [100]$ or $[110]$ measurements. It also provides a figure of merit for quantifying the hysteresis of the MMPT: upon cooling the sample, the extracted RMS deviation of the magnetic-field maps shows a similar behavior as along the heating branch, but the curve is shifted ~20 K lower. This value agrees well with the hysteresis width extracted from the temperature-dependent vibrating-sample magnetometry measurements (Fig. 1b).

Magnetic imaging and statistical analysis of the metamagnetic phase transition of FeRh with electron spins in diamond

*2. 3 Statistical analysis of the anisotropy*

Recent numerical studies have concluded that the magnetocrystalline anisotropy (MCA) of FeRh thin films depends on epitaxial strain[21,22]. Indeed, samples fabricated under different experimental conditions have been found to exhibit different degrees of MCA[22–24]. We analyze the magnetocrystalline anisotropy of the sample using the two-dimensional autocorrelation function

$$R_{\mathrm{BB}} = \sum_{x,y} B_{\mathrm{NV}}(x + \delta x, y + \delta y)B_{\mathrm{NV}}(x, y), \quad (2)$$

where we sum over each pixel, identified by its coordinates $(x, y)$, and $\delta x$ and $\delta y$ are displacements in the $\hat{x}$ and $\hat{y}$ directions. This quantity gauges how similar the field at $(x, y)$ is to that at $(x + \delta x, y + \delta y)$ and is therefore a useful tool for quantifying the characteristic length scales of a magnetic-field map.

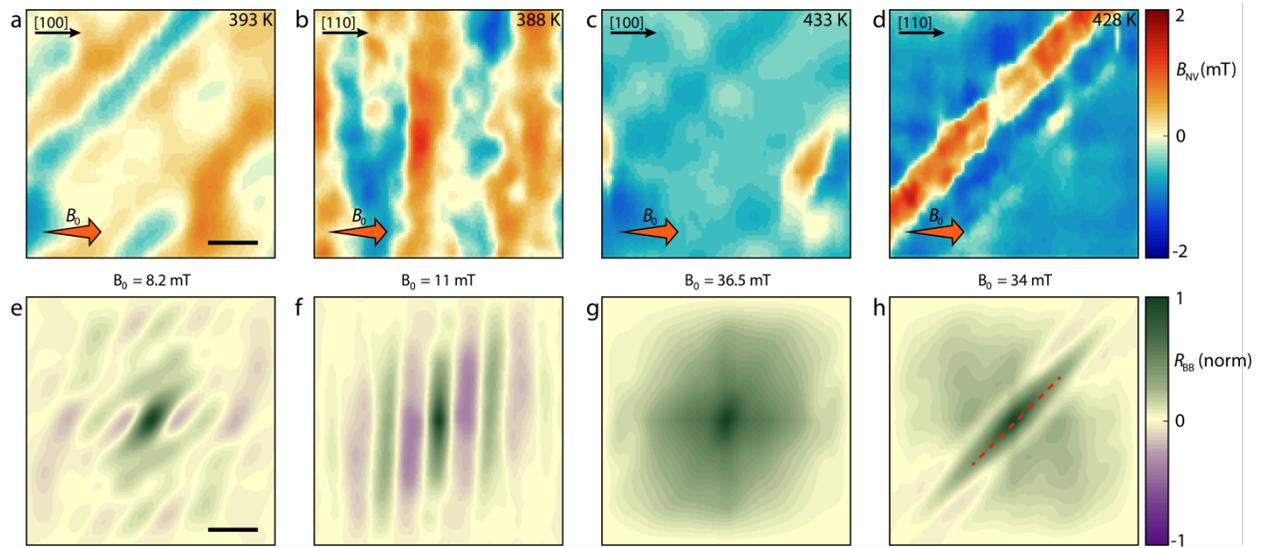

**Figure 3. Statistical characterization of the magnetocrystalline anisotropy of FeRh. (a) – (d)** Stray-field maps for small bias field and temperature moderately above $T_t$ (a-b), and large bias field and temperature well above $T_t$ (c-d). In (a), (c) the in-plane component of the bias field is parallel to an easy axis, while in (b), (d) it is parallel to a hard axis. Black arrows indicate crystalline directions of the FeRh lattice. Scale bar, 10 μm. **(e) – (h)** Normalized two-dimensional autocorrelation maps corresponding to the magnetic field distributions in (a) – (d), respectively. The scarlet dashed line in (h) indicates the FWHM of the central peak of the autocorrelation map, measured along [100]. Scale bar, 20 μm.

For a relatively low bias field (~10 mT) and a temperature ~20 K above $T_t$ (Fig. 3a – b), the measured $B_{\mathrm{NV}}$ maps display patterns with diagonal or vertical stripe-like features, when $\mathbf{B}_0^{\parallel}$ is aligned with a magnetic easy or hard axis of the material, respectively. These directional features are reflected in the corresponding autocorrelation maps (Fig. 3e – f). Note that the vertical and diagonal stripes in Fig. 3a – b (and equivalently Fig. 3e – f) reflect a similarly oriented domain

Magnetic imaging and statistical analysis of the metamagnetic phase transition of FeRh with electron spins in diamond structure since the edges of the maps in Fig. 3a and 3e are parallel to the ⟨100⟩ directions, while those of the maps in Fig. 3b and 3f are parallel to the ⟨110⟩ directions.

For temperatures well above (~40 K) $T_t$ and relatively high magnetic field (~35 mT), the stray-field maps become nearly homogeneous when $\mathbf{B}_0^\parallel \parallel [100]$ (easy axis), as FM domains start to occupy areas comparable to or larger than the imaged region (Fig. 3c). This translates into a broad central peak in the associated autocorrelation map (Fig. 3g), representing the large uniformity of the stray field. On the other hand, when $\mathbf{B}_0^\parallel \parallel [110]$ (hard axis), the measured field distributions also become smoother, but with well-defined diagonal features (Fig. 3d), which are manifested in the autocorrelation maps as sharp central peaks elongated along a magnetic easy axis (Fig. 3h). This result agrees with the energy competition discussed in section 2.2 and highlights how the statistical approach strengthens the interpretation of real-space images.

*2.4 Magnetic-field dependence of the MMPT*

Next, we investigate the effect of the magnitude of the bias field on the magnetic configuration and show how the autocorrelation function enables a systematic study of the size and (re)orientation of FM domains. We find that while changing the magnitude of $\mathbf{B}_0^\parallel \parallel [110]$ does not significantly affect the RMS deviation (Fig. 4a), it does alter the length scale of the spatial correlations (Fig. 4b). We evaluate this length scale by considering the full width at half maximum FWHM$_{[100]}$ of the central peak of the autocorrelation maps taken along the [100] direction (dashed line in Fig. 3h). Since FWHM$_{[100]}$ is a measure of the typical size of the features in the stray-field maps, it provides a figure of merit that is well suited to study the evolution of the size of the magnetic domains. Below $T_t$, FWHM$_{[100]}$ is large because of the presence of extensive AFM regions. At the onset of the phase transition, FWHM$_{[100]}$ drops as small FM domains emerge and start producing highly granular stray-field configurations. This corresponds to the nucleation stage of the MMPT and is consistent with previous imaging studies of FeRh[19,24,25].

For higher temperatures, the growth and coalescence of FM domains cause an increment of FWHM$_{[100]}$ that is larger for stronger bias fields. We interpret this as a result of the dominance of the Zeeman energy over the MCA energy in this temperature regime: as the bias field increases, FM domains with magnetization having a significant projection along the bias field start to expand to lower the Zeeman energy. These domains are predominantly elongated along [100] and, therefore, lead to higher FWHM$_{[100]}$ values. This interpretation could be corroborated by determining the magnetization direction of the FM domains via e.g. scanning electron microscopy with polarization analysis[24] or X-ray photoemission electron microscopy[19]. We note that NV magnetometry does not allow an unambiguous reconstruction of the magnetization orientation from the measured stray field, as such reconstruction is an underconstrained inverse problem[2].

Interestingly, we detect a reorientation of the domain walls across the MMPT when the bias field is applied along a hard axis (Fig. 4c). This can be seen by analyzing the FWHM along different directions, which indicates the preferred direction of the spatial correlations in the stray-field

Magnetic imaging and statistical analysis of the metamagnetic phase transition of FeRh with electron spins in diamond maps. Specifically, by taking the ratio $FWHM_{[110]}/FWHM_{[100]}$, we observe domain walls that align with a hard axis from the start of the MMPT up to $T \approx 388$ K, as seen in Fig. 3b. Above this temperature, domain walls parallel to an easy axis start to become more recurrent. When the phase transition is completed, the orientation parallel to [100] is dominant in the case of relatively high $B_0$ (~35 mT), as illustrated in Fig. 3d.

The domain-wall realignment is concomitant with the sudden decrease of the RMS deviation (Fig. 2h). This decrease indicates a reorientation of the FM domains into a state that decreases the stray field. Thus, our results suggest that a reorientation of the FM domains is behind the domain-wall realignment. Such a rearrangement results from the changing balance between the anisotropy and Zeeman energies and could also be affected by the exchange coupling between FM and AFM regions that pins the FM domains in the first stages of the MMPT[13].

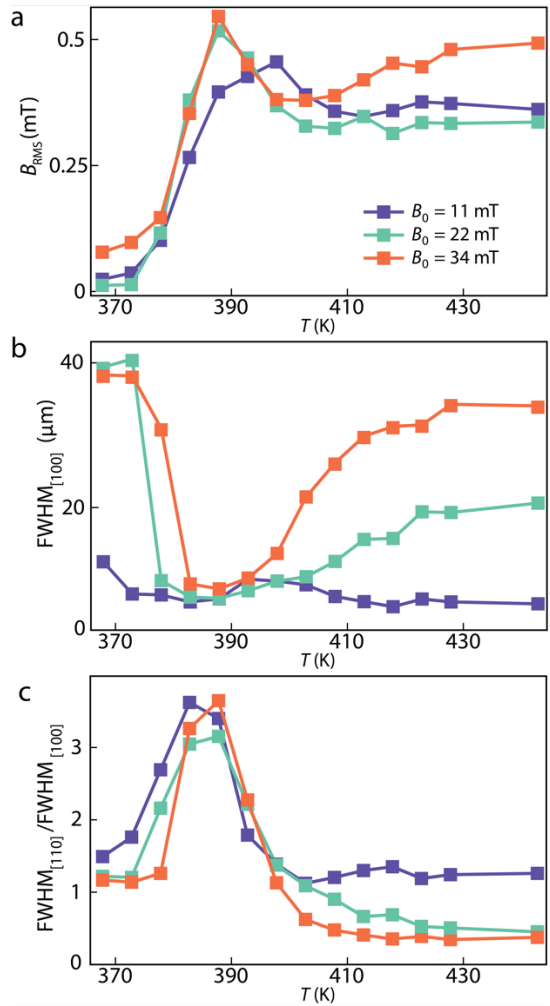

**Figure 4. Magnetic-field dependence of the metamagnetic phase transition of FeRh. (a)** RMS deviation of the measured stray field distributions as a function of temperature, for different values of the bias field along a hard axis. Each series of measurements is taken at a fixed bias field in the heating branch of the experiment. **(b)** Temperature dependence of the FWHM, measured along the [100] direction, of the central peak of the autocorrelation maps for the same values of the bias field as in (a). **(c)** Analysis of the Magnetic imaging and statistical analysis of the metamagnetic phase transition of FeRh with electron spins in diamond

rotation of domain walls in the magnetic texture of FeRh across the phase transition done by calculating the ratio of the FWHM of the central peak along [110] to the FWHM along [100], for the same bias field values of (a) and (b).

Our analysis demonstrates how the autocorrelation function and its width in different directions provide figures of merit that are well suited for tracking the evolution and orientation of magnetic domain walls as a function of control parameters (here, the bias field strength and its direction).

## 4. Conclusions

Our results show how imaging the local state of a system and quantifying the corresponding global behavior using statistical tools yield complementary insight into the temperature- and bias-field dependence of phase transitions in a magnetic material. In particular, we extracted the transition temperature and hysteresis width of the MMPT in an FeRh thin film by means of a statistical analysis of spatial stray-field images. Furthermore, we characterized the spatial correlations of the measured stray-field maps and proposed a simple way of gauging the alignment and reorientation of ferromagnetic domain walls, based on quantifying the FWHM of the autocorrelation function. The statistical methods employed here could be extended to characterize spatial correlations of stray fields generated by e.g. coherent[26] and incoherent spin-wave excitations[26–28]. The methods could also be implemented in scanning-probe NV magnetometers to gain nanometric spatial resolution[29]. This would enable studying spatial correlations and phase separation and coexistence at the nanoscale, paving the way for the imaging and characterization of phase transitions in e.g. Kagome lattices[30], spin glasses[31] and monolayer magnets[32].

**Acknowledgements**

The authors thank Sheena K.K. Patel, Eric E. Fullerton, and Jan Aarts for their support, and Ray Descoteaux for technical assistance. T.v.d.S. acknowledges funding from the Dutch Research Council (NWO) as part of the Frontiers of Nanoscience (NanoFront) program through NWO Projectruimte grant 680.91.115. The work at CMRR, UCSD was funded by the DOE grant No. DE-SC0003678.

**Data availability statement**

All data contained in the figures will be available upon publication at Zenodo.org with the identifier 10.5281/zenodo.4638723. Additional data related to this paper may be requested from the authors.

Magnetic imaging and statistical analysis of the metamagnetic phase transition of FeRh with electron spins in diamond